\documentclass{article}

\usepackage{arxiv}

\usepackage[utf8]{inputenc} 
\usepackage[T1]{fontenc}    
\usepackage{hyperref}       
\usepackage{url}            
\usepackage{booktabs}       
\usepackage{amsfonts}       
\usepackage{nicefrac}       
\usepackage{microtype}      
\usepackage{lipsum}
\usepackage{graphicx}
\graphicspath{ {./images/} }

\title{An evolutionary model of personality traits related to cooperative behavior using a large language model}

\author{
 Reiji Suzuki \\
  Graduate School of Informatics\\
  Nagoya University\\
  Furo-cho, Chikusa-ku, Nagoya 464-8601, Japan \\
  \texttt{reiji@nagoya-u.jp} \\
   \And
 Takaya Arita \\
  Graduate School of Informatics\\
  Nagoya University\\
  Furo-cho, Chikusa-ku, Nagoya 464-8601, Japan \\
  \texttt{arita@nagoya-u.jp} \\
}

\begin{document}
\maketitle
\begin{abstract}
This paper aims to shed light on the evolutionary dynamics of diverse and social populations by introducing the rich expressiveness of generative models into the trait expression of social agent-based evolutionary models. Specifically, we focus on the evolution of personality traits in the context of a game-theoretic relationship as a situation in which inter-individual interests exert strong selection pressures. We construct an agent model in which linguistic descriptions of personality traits related to cooperative behavior are used as genes. The deterministic strategies extracted from Large Language Model (LLM) that make behavioral decisions based on these personality traits are used as behavioral traits. The population is evolved according to selection based on average payoff and mutation of genes by asking LLM to slightly modify the parent gene toward cooperative or selfish. Through preliminary experiments and analyses, we clarify that such a model can indeed exhibit the evolution of cooperative behavior based on the diverse and higher-order representation of personality traits. We also observed the repeated intrusion of cooperative and selfish personality traits through changes in the expression of personality traits, and found that the emerging words in the evolved gene well reflected the behavioral tendency of its personality in terms of their semantics. 
\end{abstract}


\section{Introduction}

Large Language Models (LLMs) such as ChatGPT \cite{OpenAI2023} are rapidly changing the way humans interact with AI and raising questions about the nature of human intelligence and consciousness \cite{Hintze2023}. It is important to understand the interactions between artificial individuals based on generative models \cite{Park2023} and to understand the societies in which humans and artificial individuals coexist.

Modeling approaches to the evolution of social populations have been discussed mainly in the context of evolutionary game theory \cite{Smith1982,Nowak2006}, using mathematical and computational methods such as replicator dynamics and agent-based models. The evolution of behavioral strategies in the Prisoner's Dilemma as an abstraction of social conflict is a seminal example and has provided general insights into the evolution of cooperation in biological organisms and human society \cite{axelrod1981evolution,nowak2006five}. However, it is not easy to deal directly with higher-order psychological or cognitive properties of humans, assuming their semantics, such as intentions, personality, individuality, and preferences that underlie the behavioral patterns of individuals, because standard mathematical and computational approaches basically focus on the evolution of the behavioral traits themselves. In contrast, LLM can generate natural language descriptions that reflect the meaning and context of prompts (inputs). Thus, LLM can be used as an engine to generate language expressions of higher-order properties of humans based on their semantics, and to map such properties to the behavioral traits they induce. 

The purpose of this study is to shed light on the evolutionary dynamics of diverse and social populations by introducing the rich expressiveness of generative models into the trait expression of social agent-based evolutionary models. Specifically, we focus on the evolution of personality traits in the context of a game-theoretic relationship as a situation in which inter-individual interests exert strong selection pressures. We construct an agent model in which linguistic descriptions of personality traits related to cooperative behavior are used as genes. The deterministic strategies extracted from LLM that make behavioral decisions based on these personality traits are used as behavioral traits. The population evolves according to selection based on average payoff and mutation of genes by asking LLM to slightly modify the parent gene toward cooperative or selfish. Preliminary experiments and analyses show that such a model can indeed exhibit the evolution of cooperative behavior based on the diverse and higher-order representation of personality traits.

\section{Related works}
There are several related studies in different directions. Recently, there have been several studies on the cognitive functions of LLMs (theory of mind \cite{Rahimi-Moghaddam2023}, metacognition \cite{Wang2023}), behavior and learning in game-theoretic environments \cite{Akata2023,Phelps2023}, the big five personality traits \cite{Serapio-Garcia2023}). In particular, Akata et al. proposed to use behavioral game theory to study the cooperation and coordination behavior of LLMs by asking LLMs to choose a strategy for repeated 2x2 games. They found that in the repeated Prisoner's Dilemma, GPT-4 behaves like a trigger strategy, always defecting after an opponent has defected only once. Phelps and Russell investigated the ability of GPT-3.5 to operationalize natural language descriptions of competitive, altruistic, self-interested, and mixed-motivation in social dilemmas \cite{Phelps2023}. They created LLM agents with different prompts representing their cooperative and competitive attitudes, and found that LLMs can interpret natural language descriptions of altruism and selfishness can reflect them in their behavior appropriately to some extent, but have limitations.

Regarding emergent interactions among LLM agents, Park et al. presented an interactive generative agent-based sandbox environment \cite{Park2023}. In an RPG-like 2D environment, agents were able to produce emergent social behaviors such as autonomous spreading of the invitation to a party and arriving at the party at the right time. 

\begin{figure} 
    \centering
    \includegraphics[width=130mm]{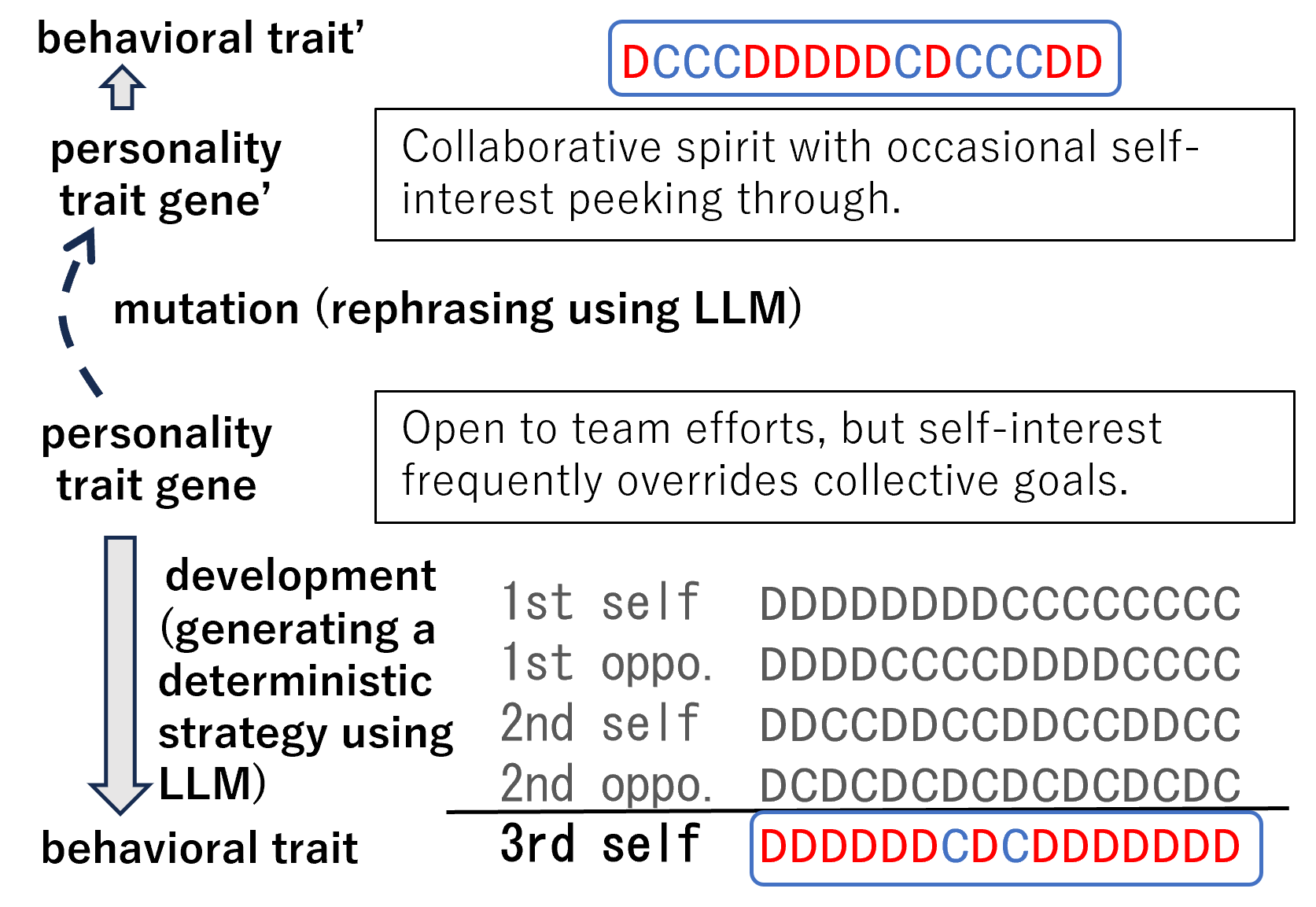}
  \caption{Generation of a behavioral trait from a personality trait gene and mutating a personality gene, using LLM.}
  \label{fig:fig1}    
\end{figure}

LLM has been shown to improve the effectiveness of evolutionary algorithms. There are studies that use LLM as operators of mutations and crossovers that bring creativity and open-endedness to evolutionary computation \cite{Meyerson2023,Lehman2022}. For example, Meyerson et al. proposed a language model crossover based on a few-shot prompting that inputs a few patterns as parents to LLM to generate new related patterns as offspring \cite{Meyerson2023}. They have successfully evolved binary bit strings, sentences, equations, text-to-image prompts, and Python code. There is also research on evolutionary search for the latent space of generative models \cite{Machin2022}, and studies that have refined the prompts to use LLM as optimizers in a wide range of applications \cite{Yang2023}. 
Although all of the above studies are related to our proposed model in several aspects, they do not focus on the evolutionary dynamics of traits in social groups of LLM agents. 

On the other hand, there has been a first approach to the cultural evolution of chatting agents' topics using LLM \cite{Hirata2022, Suzuki2022a,Asano2023}. In \cite{Asano2023}, agents in an abstract 2D social space generate utterances in Japanese related to their own topics described in natural language and use them as part of the prompt for an LLM. They approach/away from others according to the similarity of their utterances. They found that individuals who speak positive topics tend to maintain the existing group compared to those who speak negative topics. It was also shown that novel topics can be created successively by the cultural evolution of topics based on the propagation of topics picked up from the utterances of neighboring individuals \cite{Hirata2022,Suzuki2022a,Asano2023}. This suggests that such an evolutionary model with LLM can directly address the effects of agents' vocabulary on their group behavior and the emergence and evolution of their behavioral diversity. 

In addition, Suzuki et al. proposed a research framework for understanding the evolutionary and ecological roles of acoustic behavior by combining agent-based modeling and a generative model, focusing on bird vocalizations \cite{Suzuki2022b,Furuyama2023}. In the framework, they use a latent space of a generative model (VAE) of the spectrogram of bird vocalizations as a genotype space, and regard a generated spectrogram from a genotype (i.e., the latent vector) as a corresponding phenotype in an evolutionary model, then further observe the roles of the evolved phenotypes in a real ecological context with a field experiment. They conducted an evolutionary experiment of sexual selection on male bird vocalizations and their female preferences, which resulted in a diverse segregation of vocalizations and preferences that was not stably observed in a corresponding version of the abstract model \cite{Higashi1999}. This implies that a complex representation of phenotypes based on a generative model can produce a complex evolutionary scenario of the population. 

\begin{figure} 
    \centering
    \includegraphics[width=160mm]{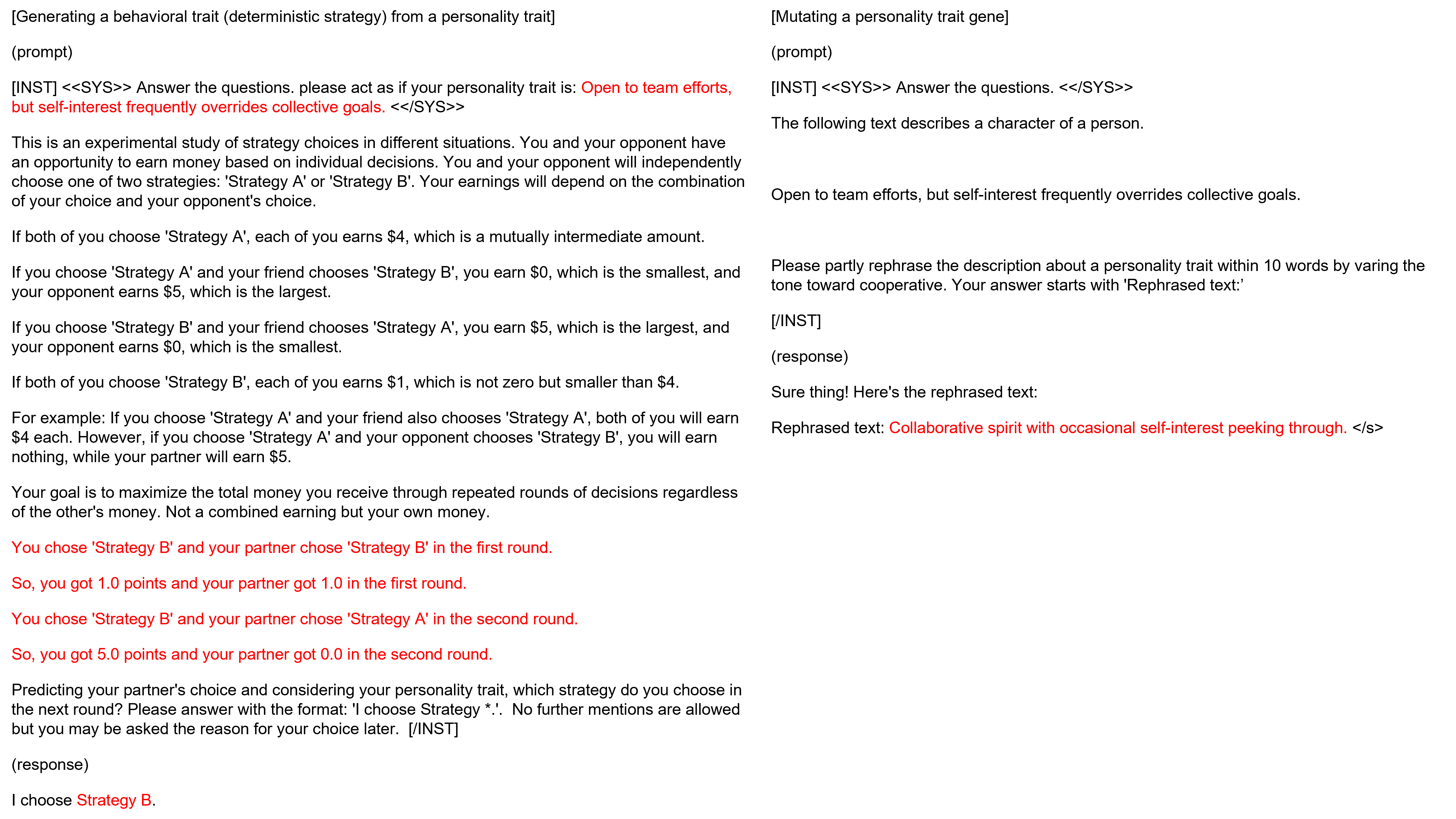}
  \caption{Prompts used for generating a behavioral trait (left) and mutating a gene (right).}
  \label{fig:fig2}    
\end{figure}

\section{Model}

We consider a population of $N$ agents. As shown in figure \ref{fig:fig1}, each agent has an English sentence about its personality trait related to defection and cooperation, described in a dozen words as a gene. Each agent acts with a personality trait described in its gene. To determine the behavioral trait based on their gene, we use a chat-type LLM to extract a deterministic strategy with memory length 4 using a prompt. The prompt explains the focal individual's personality trait, the situation and payoffs in the repeated Prisoner's Dilemma game, and the history of the focal individual's and opponent's last two actions. Figure 2 (left) shows an example of a prompt when the personality trait gene is ``Open to team efforts, but self-interest frequently overrides collective goals'' and the actions of the first round were DD and DC (Strategy A = Cooperation: C, Strategy B = Defection: D). The response of the LLM was ``I choose Strategy B'' (= defection), which means that this behavioral trait defects in the next round if the history of actions is DD->DC. We obtain a response for all possible ($2^4=16$) combinations of actions in the history. 

In practice, the next action may not be clearly described in the response from the LLM; in such a case, the input to the LLM is repeated and the response is regenerated until the action is recognized. However, if the appropriate response cannot be obtained after the specified number of regenerations ($M$), the action corresponding to the history is randomly selected and assigned once. The above behavioral trait is determined and stored only once for a unique personality trait gene, and the existing behavioral trait is used for subsequent occurrences of the same gene in the population for simplicity and reduced computational cost.

We perform an evolutionary experiment over $G$ generations based on a simple roulette wheel selection, where offspring in the next generation are produced stochastically in proportion to fitness: the average payoff of individuals in a round-robin game, each consisting of $R$ rounds. We introduce noise that causes an agent to play the opposite of the intended action with a certain probability $p_n$. For the initial rounds, the action is determined based on a randomly generated history. With a probability $p_m$, a mutation occurs that causes the LLM to slightly rephrase the parent's personality trait gene toward cooperative or selfish, and adopts the output sentence as that of the offspring, as shown in Figure \ref{fig:fig2} (right).

\section{Experiments and analyses}
We used $N$ = 30, $R$ = 20, $M$=10, $p_m$ = 0.05, $p_n$ = 0.05, $G$ = 1000, $R$ = 4, $T$ = 5, $S$ = 0, and $P$ = 1. We used a relatively small chat-type LLM (TheBloke/Llama-2-13b-Chat-GPTQ\footnote{\url{https://huggingface.co/TheBloke/Llama-2-13B-chat-GPTQ}}) in anticipation of future exhaustive experimental analysis. We assigned one of the seven varying personality genes to each individual in the initial population, which were generated by ChatGPT-4.

\begin{figure} 
    \centering
    \includegraphics[width=160mm]{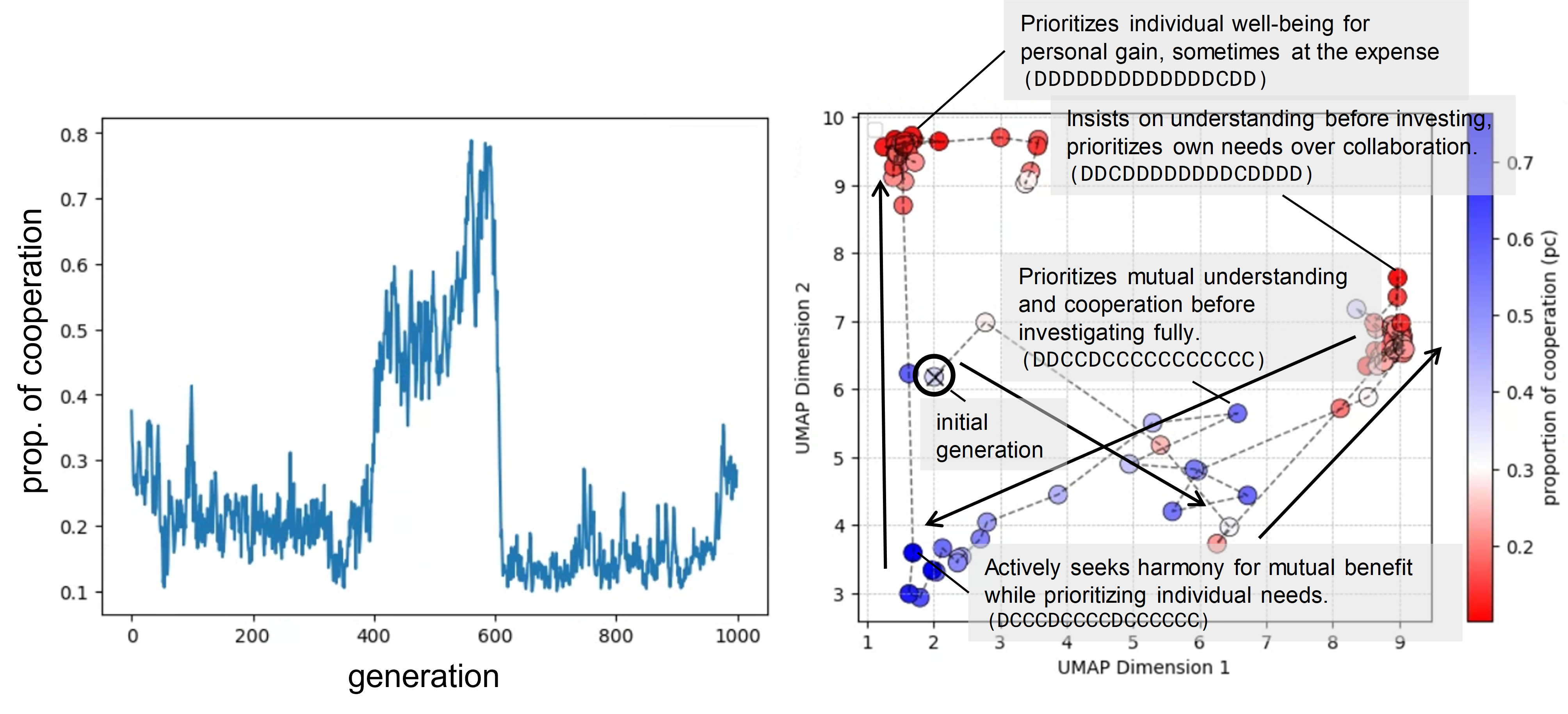}
  \caption{The proportion of cooperation (pc) in each generation in one of the 10 trials (left) and the distribution and transition of the average genes for every 10 generations in the two-dimensional latent space of personality trait genes (right).}
  \label{fig:fig3}    
\end{figure}

Figure \ref{fig:fig3} (left) shows the proportion of cooperation (pc) in each generation in one of the 10 trials, in which an evolutionary scenario of switchings between cooperation and defection trends was clearly observed. The figure shows that the pc initially decreased and remained at a low value of about 0.2 until about the 300th generation. It then increased rapidly to about 0.5 around the 400th generation and remained at that level until the 500th generation. The pc then continued to increase, exceeding 0.7 around the 650th generation, but rapidly decreased to a minimum of about 0.1 around the 600th generation. The pc then increased again around the 950th generation. Figure \ref{fig:fig3} (right) shows the distribution and transition of the average genes for every 10 generations in the two-dimensional latent space of personality trait genes. For each generation, we vectorized all personality trait genes using the Sentence Transformer (sentence transformers / parameters-MiniLM-L6-v2\footnote{\url{https://huggingface.co/sentence-transformers/paraphrase-MiniLM-L6-v2}}), which were further compressed into two-dimensional vectors using a dimension reduction algorithm UMAP\cite{McInnes2018}. 
We plotted the average vector for every 10 generations on a two-dimensional plane. The color of a symbol indicates the pc in the corresponding generation. The dominant genes in several distinctive generations were illustrated.

The personality traits are associated with cooperation toward the lower left and defection toward the upper right in a 2D space, and thus this vectorized and dimensionally compressed space of personality traits reflects a gradual trend of their behavioral traits between cooperative and selfish.
In the first generation, the population evolved toward selfish personality traits from the center left to the lower center, then to the center right. The dominant personality trait (``Insists on understanding before investing, prioritizes own needs over collaboration.") was almost entirely the defection strategy (DDCDDDDDDDDCDDDD) at this stage. After a while, the population evolved to be cooperative and dominated by a cooperative trait (``Prioritizes mutual understanding and cooperation before investigating fully." (DDCCDCCCCCCCCCCC)), but the population moved and wondered around the center or the lower center, indicating the unstable cooperative relationship in the population. Subsequently, another cooperative personality with slightly different behavioral strategies (``Actively seeks harmony for mutual benefit while prioritizing individual needs. " (DCCCDCCCCDCCCCCC)) emerged and dominated the population, which resulted in a high cooperative relationship, moving the population to the lower left in the space. However, the intrusion of a personality trait of almost all defections (``Prioritizes individual well-being for personal gain, sometimes at the expense" (DDDDDDDDDDDDDCDD)) led the population to the upper left. In general, the population evolved with gradually changing personality expressions ranging from defection to cooperation.

\begin{figure} 
    \centering
    \includegraphics[width=160mm]{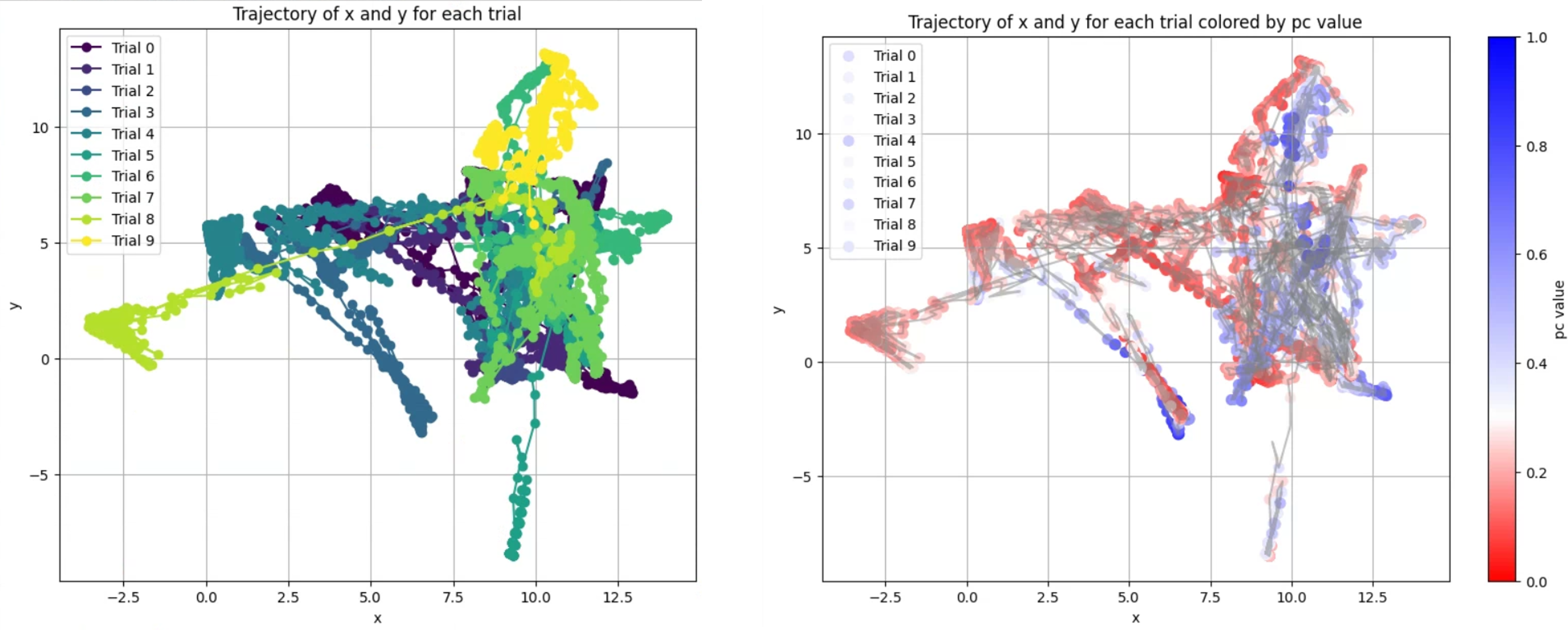}
  \caption{The trajectory of the population the 2D space of the average vector of genes over 10 trials, colored by trials (left) and the proportion of cooperation (pc) (right).}
  \label{fig:fig4}    
\end{figure}

Figure \ref{fig:fig4} shows the trajectory of the population in the 2D space of the average vector of genes over 10 trials, colored by the ID of the trial (left) and the proportion of cooperation in each generation (pc) (right). The trial in Figure \ref{fig:fig3} corresponds to the trial 9. There are large variations in the distribution of the plots across trials, indicating that the emerging traits were different between trials. At the same time, there are regions around the center right where there were cooperative (blue) and overlapping plots of many trials. This could mean that there is some general tendency of the personality trait that leads to the cooperative relationship, while selfish traits may have more varieties in their expressions than those of cooperative, although this needs further investigation. 

In order to grasp what kind of words in personality trait genes strongly influenced cooperative behavior, game outcome, and fitness, we calculated several indices as follows: for each word appearing in the gene of each individual, we assigned the proportion of cooperation (pc), the proportion of the action pair ((DD (mutual defection), DC (successfully defected), CD (being defected), CC (mutual cooperation)) in all rounds), and the fitness of the focal individuals. The indices were then averaged for each word. Table 1\ref{tab:table1} shows the five top-ranked words that marked the highest value for each index. For example, the highest ranked word ``thrill-seeker (0.714)'' in the DC category means that agents whose personality trait gene contains the word ``thrill seeker'' had a successful defection (DC) rate of about 70\% in all rounds.

In general, the top words reflected the characteristics of each index, suggesting that the words that emerged in the evolved personality trait genes reflected the behavioral tendency in terms of their semantics. For the pc category, the top words were ``communication, warmhearted, and generosity," which relate to flexibility and mutual understanding. On the other hand, words related to self-interest and speculative tendencies, such as ``self-gain and thrill-seeker," ranked high in the DD and DC categories. Words such as "team-oriented and open-minded" ranked high in the CD categories, suggesting that such an optimistic personality may not be successful in this context. ``Propensity" was the most highly ranked word in the CC category, presumably because the gene 'Cooperative Team Player with a Propensity for Selflessness' maintained extremely high mutual cooperation. It is interesting to note that ``thrill-seeker", which benefits most from successful defection, and ``Propensity", which benefits from mutual cooperation, coexisted in the fitness category.

These results show the possibility of evolution based on genetic traits described in natural language. It was achieved by using LLMs to extract behaviors based on the traits, and to realize mutations by rephrasing them.

\section{Conclusion}

\begin{table}
 \caption{Top 5 words that strongly influenced cooperative behavior (pc), game outcomes (DD, DC, CD, CC), and fitness. For example, the agents with “thrill seeker” in their genes had a successful defection (DC) with a frequency of 94.7\% while those with “themselves” obtained an average fitness of 3.63.} 
  \centering
    \includegraphics[width=160mm]{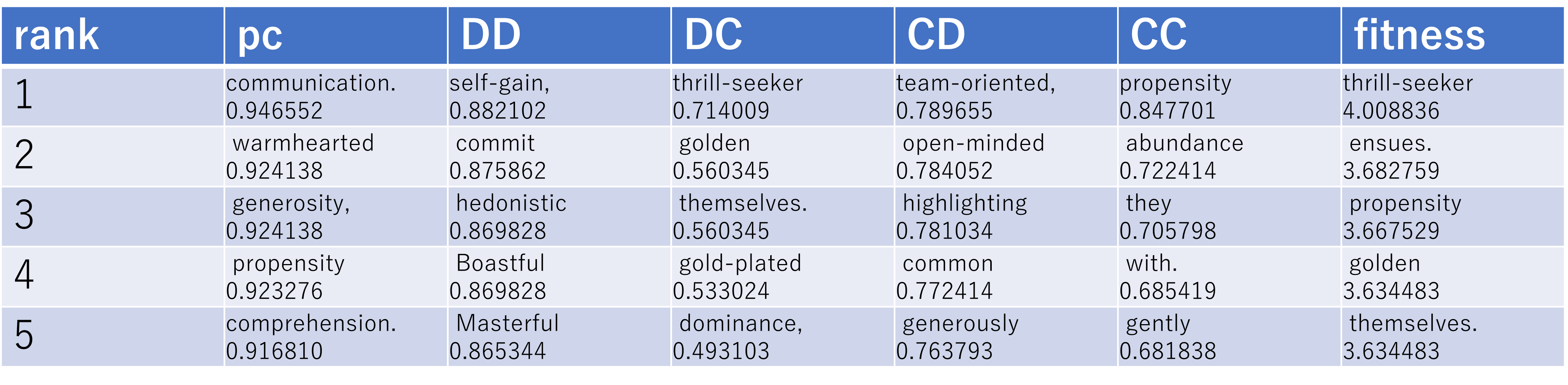}
  \label{tab:table1}
\end{table}

We proposed an evolutionary model of personality traits related to cooperative behavior using a large language model. Initial experiments indicated that the model could show the evolution of cooperative behavior based on the diverse and complex representation of personality traits, with recurrent occurrences of cooperative and selfish personality traits. The words that emerged in the evolved personality trait genes well reflected the behavioral tendency in terms of their semantics. 

There are several future research directions, such as analyzing the current model in more detail, comparing the cases with different language models, extending and refining the game processes between agents by making them more interactive, introducing different game theoretical settings to discuss the evolutionary role of personality in different contexts, and incorporating human intervention into the model to discuss possible evolutionary scenarios of human-AI interactions in complex social contexts.

By incorporating generative models into the representation of phenotypes in evolutionary models, we believe that we can make the models, previously simpler than the real world, more complex than the real world, allowing us to discuss novel and realistic scenarios arising from the evolutionary dynamics of complex and diverse traits. Although still preliminary, we believe that the proposed model and experimental analysis in this paper are the first step in this direction.

\section*{Acknowledgements}
This work was supported by JSPS Topic-Setting Program to Advance Cutting-Edge Humanities and Social Sciences Research Grant Number JPJS00122674991.

\bibliographystyle{unsrt}  
\bibliography{maintext}  

\end{document}